\documentclass{article}
\usepackage[utf8]{inputenc}
\usepackage{authblk}
\usepackage{comment} 
\usepackage{sidecap}
\usepackage{amsmath}
\usepackage{graphicx}
\usepackage{lmodern}
\usepackage[T1]{fontenc}
\usepackage{tgtermes}
\usepackage{lipsum}
\usepackage[a4paper, total={6in, 8in}]{geometry}
\usepackage{hyperref,graphicx}   
\usepackage{lineno}
 
\title{Multi-scale Analysis of Nitrogen Loss Mitigation in the US Corn Belt}
\author[1,\*]{Jing Liu\thanks{Corresponding author, email: liu207@purdue.edu}}
\author[1]{Laura Bowling}
\author[2]{Christopher Kucharik}
\author[1]{Sadia Jame}
\author[1]{Uris Baldos}
\author[3]{Larissa Jarvis} 
\author[4]{Navin Ramankutty} 
\author[1]{Thomas Hertel} 

\affil[1]{Purdue University}
\affil[2]{University of Wisconsin–Madison}
\affil[3]{McGill University}
\affil[4]{University of British Columbia}

\date{\today}

\begin{document}

\maketitle
\begin{abstract}
Reducing the size of the hypoxic zone in the Gulf of Mexico has proven to be a challenging task. A variety of mitigation options have been proposed, each likely to produce markedly different patterns of mitigation with widely varying consequences for the economy. The general consensus is that no single measure alone is sufficient to achieve the EPA Task Force goal for reducing the Gulf hypoxic zone and it appears that a combination of management practices must be employed. However, absent a highly resolved, multi-scale framework for assessing these policy combinations, it has been unclear what pattern of mitigation is likely to emerge from different policies and what the consequences would be for local, regional and national land use, food prices and farm returns. We address this research gap by utilizing a novel multi-scale framework for evaluating alternative N loss management policies in the Mississippi River basin. This combines fine-scale agro-ecosystem responses with an economic model capturing domestic and international market and price linkages. We find that wetland restoration combined with improved N use efficiency, along with a leaching tax could reduce the Mississippi River N load by 30-53\% while only modestly increasing corn prices. This study underscores the value of fine-resolution analysis and the potential of combined economic and ecological instruments in tackling nonpoint source nitrate pollution. 
\end{abstract}

\section*{Introduction}

Widespread and intensive agricultural activity has resulted in large amounts of nitrogen (N) leaching from soils \cite{turner_emergence_2007-1,goolsby_nitrogen_2001}. Elevated N levels in streams and rivers causes a spectrum of challenging problems including biodiversity loss and threatened human health \cite{vitousek_human_1997}. Nutrients transported through the Mississippi River basin have been blamed for what are referred to as the "dead zones" (low oxygen water) formed in the Gulf of Mexico \cite{diaz_spreading_2008, rabalais_hypoxia_2001}. The largest hypoxic zone measured since 1985 was 8,776 square miles in 2017 (US-EPA Hypoxia Task Force). Reducing this size to an acceptable level by 2035 will require at least a 45\% reduction in the N load exported by the Mississippi and Atchafalaya Rivers \cite{USEPA2013, scavia2017ensemble}.

It is widely recognized that there is no silver bullet for resolving the wicked problem of nonpoint source water pollution in the Mississippi watershed \cite{mclellan_right_2018}. To achieve the 45\% nutrient reduction goal, in-field nutrient management must be combined with edge of field measures as well as downstream nutrient removal practices \cite{mclellan_right_2018, schilling_modeling_2009,iowa2013science, mclellan_reducing_2015}. While agronomic and environmental management techniques to control and remove lost N have advanced, there is limited evidence that existing policies are effective in facilitating the adoption of these techniques \cite{shortle_reforming_2012, mclellan2018right, roy2021hot}. The programs to promote improved water quality in the US have been found to be largely inefficient as the incremental cost for water quality protection has exceeded the incremental benefits \cite{olmstead_economics_2010, shortle_reforming_2012,laukkanen_evaluating_2014}. The low efficiency is often attributed to the failure to identify the proper value of N effluent mitigation. The uniform value assumed in current policy design does not reflect the spatially varying marginal cost of mitigating water quality damages \cite{shortle_nutrient_2017}. Quantifying this cost is challenging in practice because nonpoint source pollution is often not measurable. Without knowing the site-specific biophysical and ecological characteristics of N leaching, economic instruments cannot be efficiently deployed. 

Our paper overcomes these problems by estimating key biophysical relationships between corn production, N fertilizer use, and nitrate pollution based on the fine-scale, agro-ecosystem model Agro-IBIS \cite{kucharik_evaluation_2003, kucharik_integrated_2003, donner_corn-based_2008}. These biophysical relationships are built into a grid-resolving economic model dubbed SIMPLE-G-US-CS. This model embeds a fine-scale analysis of the continental US within a global economic model, thereby allowing for a linkage to be established between grid-cell specific interventions and national and international markets for corn and soybeans, as well as N fertilizer and other inputs. Policy-induced, market-mediated spillovers are found to be an important part of the economic consequences of nutrient mitigation strategies. More details about the SIMPLE-G-US-CS model can be found in the Materials and Methods section as well as in the Supporting Information Appendix. Using this integrated framework, we compare the effectiveness of various policies to reduce nitrate loading in the Mississippi River basin and the spatial patterns of mitigation. Four individual policies as well as various policy combinations are examined, including: (A) an N leaching tax, (B) improved N use efficiency, (C) controlled drainage (the use of adjustable, flow-retarding structures placed in the drainage system that allow the water depth to be adjusted), and (D) wetland restoration. Among them, policies A\&B are nationwide, while C\&D are constrained by the feasibility of the practices \cite{jaynes2010potential, cheng2020maximizing}.

\section*{Results}
\subsection*{No single mitigation strategy alone is sufficient to achieve the Hypoxia Task Force goal of reducing the N loading in the Gulf of Mexico by 45\%} 
\begin{figure}[tbhp]
\centering
\includegraphics[width=.95\linewidth]{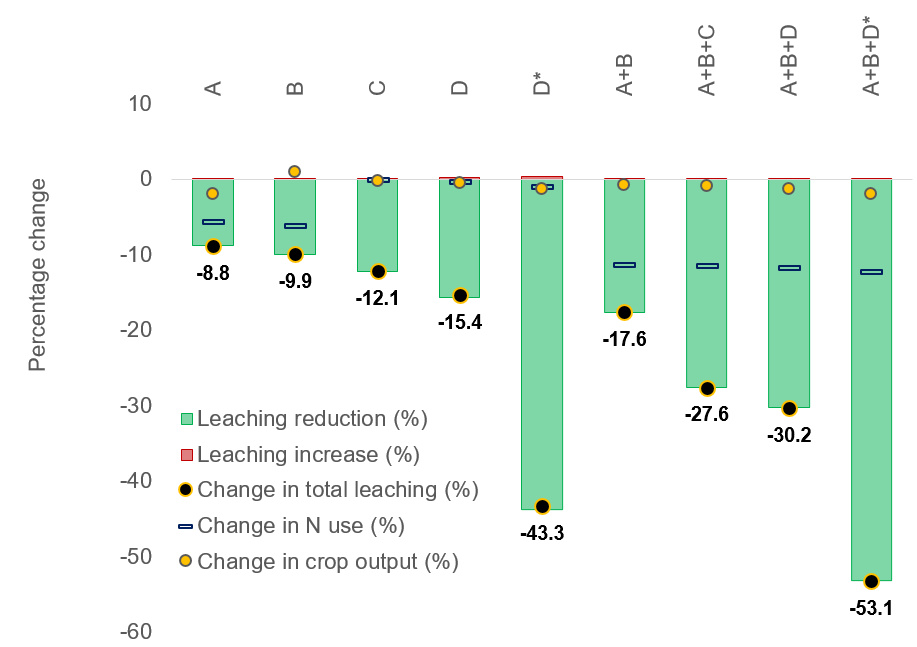}
\caption{Changes in nitrate leaching, N fertilizer use, and crop output (in \%) associated with strategy A (a leaching tax), strategy B (improved N use efficiency), strategy C (controlled drainage), strategy D and D* (wetland restoration), and the combination of these strategies. Strategy D assumes wetland is constructed only where tile drainage is already installed, while strategy D* is not subject to this constraint.}  

\label{fig:f1}
\end{figure}

Among the four policies explored, wetland restoration appears to be most effective, reducing leaching by 15-43\% (Figure \ref{fig:f1}). A tax rate of one dollar per ton of leaching (strategy A) that boosts the average cost of N fertilizer to corn farms by 28.9\% reduces N fertilizer use by 5.7\% and total leaching by 8.8\%. Improving N use efficiency by 10\% through enhanced placement and fall-spring split application, which increases N fertilizer productivity in our model (strategy B), reduces leaching from corn production within the Mississippi River basin by 9.9\%. Controlled drainage (strategy C) has a stronger impact overall, reducing N leaching by 12\%. However, when combined, wetland restoration, along with a leaching tax and increased N use efficiency (strategy A+B+D) can achieve nearly two-thirds of the 45\% reduction goal. This would represent a great improvement over the current situation. Furthermore, depending on the feasibility of establishing effective mitigation wetlands in the absence of tile drainage, this reduction could be much larger (43\% alone in strategy D* and 53\% combined in strategy A+B+D*). 

Crop output falls in almost all cases because of the higher input cost associated with the rising N fertilizer price in strategy A, control system installation and maintenance costs in strategy C, and wetland construction and foregone cropland in strategy D. The exception is strategy B where more efficient fertilizer use reduces the application by 6.1\% while slightly increasing crop output. The composite corn-soy price increases by no more than 3\% regardless of the scenarios due to the modest change in crop output. The sum of the individual scenarios is greater than when they are implemented in concert, indicating the presence of interactions.

\subsection*{The spatial pattern of N leaching reduction varies across strategies} 
\begin{figure}[tbhp!]
\centering
\includegraphics[width=1\linewidth]{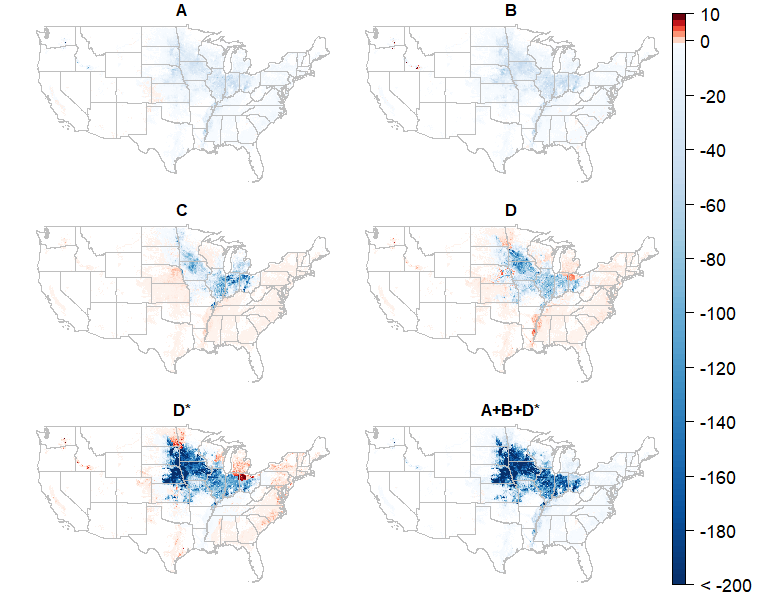}
\caption{Change of N leaching by grid-cell under strategies A, B, C, D, D* and A+B+D* (in metric tons per grid-cell). Negative value indicates leaching reduction.}
\label{fig:f2}
\end{figure}

Although total mitigation across the four individual strategies considered here is comparable, the spatial pattern of the leaching reductions can vary widely depending on the policy (Figure \ref{fig:f2}). The amount of mitigation is relatively consistent over the US Corn Belt under strategies A and B, as these are both policies that are not tied to specific locations. Such a feature also limits the market-mediated leakage. This is in marked contrast to the patterns associated with strategies C and D, policies that are contingent on local conditions \cite{jaynes2010potential, cheng2020maximizing}. Controlled drainage policies are only possible in locations where tile drainage is currently employed. And similarly, in our analysis, wetland restoration is also limited to locations where hydric soils are present. Therefore, while aggregate mitigation is comparable across these four strategies, N leaching mitigation per grid cell is much higher in the heart of the Corn Belt where controlled drainage and wetland restoration are more prevalent. These two policies also exhibit much higher N removal rates in general, compared to the first two strategies \cite{iowa2013science}.

However, conservation systems like controlled drainage and wetland restoration require farmers to incur additional costs that reduce profitability and curb output on adopting farms\footnote{We assume that farmers are responsible for half of the costs for controlled drainage and wetland installation and maintenance.}. Furthermore, wetland restoration removes land from production, although some lands are intentionally retired due to their low productivity\footnote{Following Marshall et al. \cite{marshall2018reducing} and Christianson et al. \cite{christianson2013financial}, we assume that treating 100 acres of cropland area in strategy D* (or 100 acres of tile-drained cropland land in strategy D) requires 0.5 acre of restored wetland plus 1.75 acres of wetland buffer.}. As a result, output on the adopting farms is likely to be reduced, thereby boosting corn prices, leading to output expansion and potentially additional N leaching elsewhere. This leakage effect can be seen from Figure \ref{fig:f2} where leaching around the fringes of the Corn Belt rises under strategies C, D and D* in response to higher corn prices. Because there is little tile drainage in these fringe areas, less of this increased leaching will contribute directly to the hypoxia problem in the Gulf of Mexico, but it could result in groundwater contamination. Nevertheless, the amount of reduction strongly outweighs increases in less intensive cropping areas. By combining policy D* with A and B, it is also possible to eliminate the policy spillovers that arise when either D* is implemented alone (Figure \ref{fig:f2}, policy D* vs. A+B+D*). 

\subsection*{The most effective leaching reduction policy varies by location and by state} 
\begin{figure}[tbhp!]
\centering
\includegraphics[width=1 \linewidth]{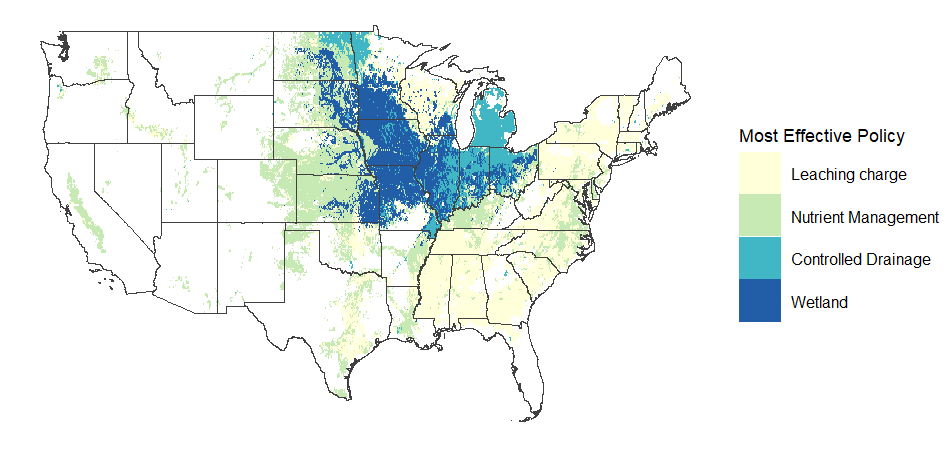}
\caption{The most effective single policy that yields the largest N leaching reduction (in terms of tons per grid-cell) identified at the grid-cell level. }
\label{fig:f3}
\end{figure}

The gridded results reported in Figure \ref{fig:f2} allow us to identify the single practice among the four that yields the largest leaching reduction (in terms of tons) at each location. This is depicted in Figure \ref{fig:f3}, which shows that controlled drainage and wetland restoration are the most effective policies across most of the Corn Belt, except for the western edge where improved N use efficiency is more effective. Outside of the Corn Belt, the leaching tax stands out as most effective, especially in the Eastern US. This stems from a combination of factors, including high N leaching intensity, reduced prevalence of tile drainage and relatively lower marginal productivity of N applications. (See Figures \ref{fig:fs-2}  and \ref{fig:fs-3} in SI Appendix.) 

\begin{figure}[tbhp]
\centering
\includegraphics[width=1\linewidth]{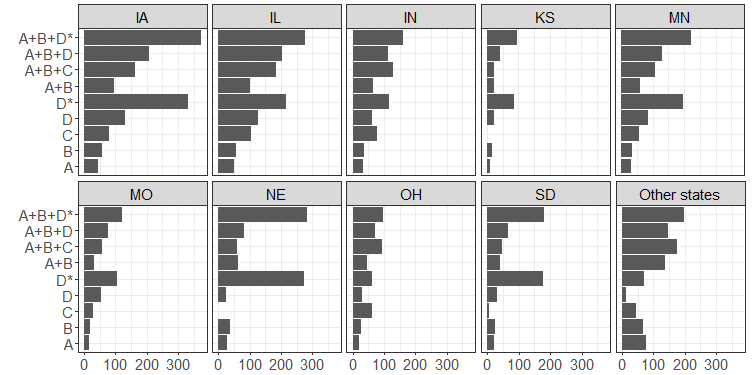}
\caption{N leaching reduction by state and by mitigation strategy. X axis shows N leaching reduction by state (in 1000 metric tons).}
\label{fig:f4}
\end{figure}

Since environmental policies are typically set at the state or federal level, not at the level of individual grid cells, we also report state level mitigation potential for the four strategies, as well as combined approaches in Figure \ref{fig:f4}. The top ten states that produce 80\% of the corn and soy output in the US use 83\% of N fertilizer and account for 80-85\% of total leaching reduction under strategies A and B, and almost all leaching reduction in strategies C and D. Controlled drainage is quite effective for Iowa, Illinois, Indiana, Minnesota,  and Ohio where tile drainage is extensively used \cite{valayamkunnath2020mapping}. Wetland restoration is also quite effective in most of the Corn Belt. However, due to the presence of market-mediated spillover effects, these policies result in higher N leaching rates in those states without potential for controlled drainage and wetland systems. Not surprisingly, policy combinations can dominate individual policies, and this difference is particularly pronounced in Iowa, Illinois, and Minnesota. 

\subsection*{Half of the total reduction in leaching originates from a small set of locations. Targeting hotspots of N leaching would make N leaching mitigation policies more efficient} Substantial reductions of N leaching are spatially concentrated in locations with large corn acreage, intensive N fertilizer use, and/or highly effective conservation practices. We find that, across all four practices, 50\% of the mitigation is contributed by just 10\% or fewer of the (total 48,317) grid cells, and the associated crop output reduction is small (5\% or less) (Figure \ref{fig:f5}). The locations of these top-mitigating grid-cells are shown in the supplementary materials (Figure \ref{fig:fs-9}). While representing just 10\% of the grid-cells, these top-mitigating locations cover 39.4\% of corn-soy area and produce 38.8\% of the corn-soy output in the US. They account for 42.5\% of N fertilizer use and generate 46.7\% of N leaching among all US corn-soy production according to our baseline data. Mitigation in the absence of edge of field practices such as wetland restoration and controlled drainage requires more dramatic rate reductions to achieve the same level of mitigation (Figure \ref{fig:f6}). Under strategies A and B (individually or combined), cutting back N use by 10\% yields approximately 15\% of leaching reduction (filled and empty red and orange dots clustered between the 2:1 and 1:1 ratio lines in Figure \ref{fig:f6}). Wetland restoration and controlled drainage are more efficient in terms of requiring less reduction of N application per unit of N leaching mitigation achieved (empty blue and green dots clustered around the 25:1 ratio line in Figure \ref{fig:f6}).  

\begin{figure}
\includegraphics[width=\linewidth]{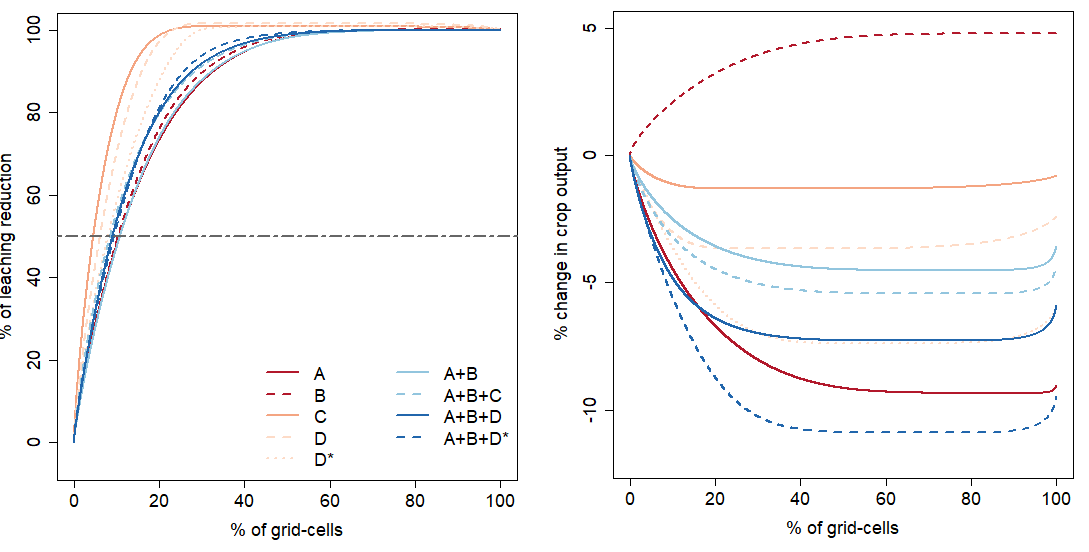}
\caption{Y axis shows the percent of total N leaching reduction (left panel) and cumulative change in crop output in percentage terms (right panel). X axis shows the percent of the number of grid cells. Grid cells are sorted descending by their level of mitigation, the order for which varies by policy. Dotted dashed horizontal line indicates 50\% of total N leaching reduction. }
\label{fig:f5}
\end{figure}

\begin{figure}[tbhp]
\centering
\includegraphics[width=\linewidth]{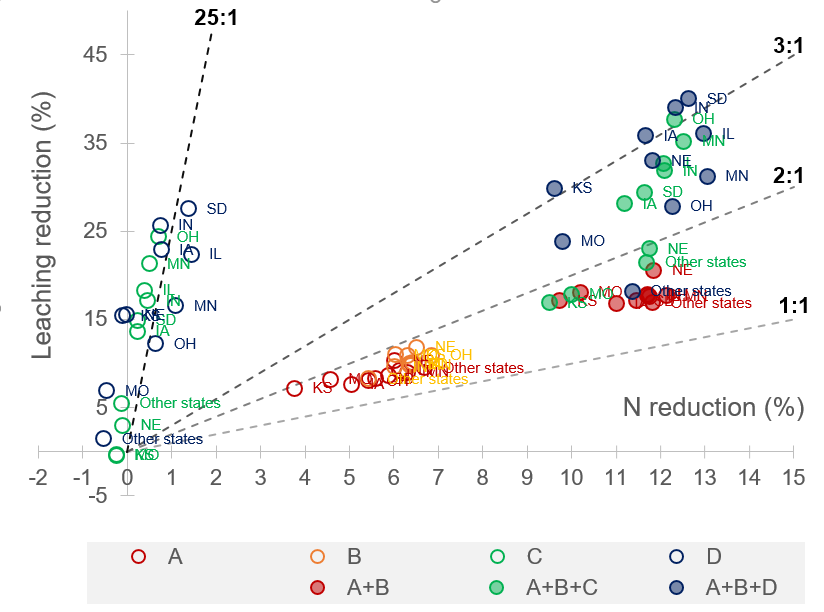}
\caption{N leaching reduction versus N use reduction (both in \%) by state. Dash lines represent the ratio of the two at 25:1, 3:1, 2:1, and 1:1. Flatter lines indicate more reduction of N application required to achieve one unit of leaching reduction, thereby potentially more disturbance to crop output.}
\label{fig:f6}
\end{figure}

\subsection*{Discussion}
The novel, multiscale framework used in this research has allowed us to identify location-specific N leaching mitigation potential as well as the market-mediated spillover effects of policies. The latter arises when the policy interventions boost corn prices while depressing N fertilizer prices, thereby inducing increased leaching in untreated grid cells.  Similar leakage effects associated with spatially targeted environmental policy intervention have been well recognized in the deployment of deforestation policies \cite{lima2019leakage, dou2018spillover} and air pollution policies \cite{fang2019clean}, but the relevant literature is sparse in the context of water pollution. We find that these spillover effects can be significant, particularly in the case of wetland restoration where cropland is removed from production. Spillover effects depend on the cost sharing between farmers and the government. The higher the farmers' burden, the greater the increase in costs and the larger the ensuing output reduction, thereby leading to greater market-mediated spillovers. Of course, the magnitude of these spillovers will depend on the specifics of the policy implementation, including the extent of wetlands, the size of surrounding buffer zones, co-payments required of farmers and adoption rates. 

Previous studies have also identified wetland restoration as a promising policy option \cite{mclellan2015reducing, fisher2004wetland}. This begs the question: Why have these not been more widely adopted? The literature suggests that many factors could affect farmers' adoption decisions. In addition to self-identification, environmental attitude, previous adoption of conservation practices, farm size, soil quality  \cite{prokopy2019adoption} and land tenure \cite{masuda2021rented}, one widely shared concern is over the adverse yield impacts from N use reductions \cite{niles2013perceptions}. A recent paper by Roy et al. \cite{roy2021hot} (county-level analysis) lends some support to this concern by showing that many Midwest counties are still below the N input break-point beyond which crop yield plateaus or declines. However, our analysis, building on the Agro-IBIS model (grid-cell level analysis), finds only a modest output response to N use reduction. The resulting national output reduction is fairly small, no more than 2\% regardless which strategy is followed. Assessing this concern requires more understanding about site-specific N balance, N breakpoints and local production technologies. 

The significant differences in mitigation patterns associated with the individual policies will have important consequences for the amount of nitrate actually reaching the Gulf due to differential N transport through the Mississippi River basin \cite{masuda2021rented}. In future work it will be important to connect this multi-scale analytical framework with hydro-ecological models capable of routing excess nutrients through the ecosystem.

\subsection*{Materials and Methods}
SIMPLE-G-US-CS is a variant of the SIMPLE-G family of models \cite{liu_achieving_2017, baldos2020simple} that has been used to study agriculture related sustainability issues. This corn-soy focused model shares many features in common with the SIMPLE-G-US model \cite{baldos2020simple} but focuses on just two crops - corn and soybeans, given that corn is an N intensive crop and often rotates with soybeans. Like the other models in the SIMPLE-G family, SIMPLE-G-US-CS is a grid-resolving global partial equilibrium model, which captures both the local response to larger-scale policy shocks as well as the aggregated local feedbacks to the regional, national, and global economies. The corn-soy model advances the general SIMPLE-G framework by introducing grid-cell level parameters estimated from an agro-ecosystem model Agro-IBIS \cite{kucharik_evaluation_2003, kucharik_miscanthus_2013, donner_corn-based_2008}. Details of the two models as well as the coupling of the two are explained in the supplementary materials.   

Four individual policies are considered in our experimental design to study the impacts of different mitigation strategies. Strategy A considers a leaching tax to increase the cost of N fertilizer application. The final cost is determined by the nationally uniform N fertilizer price (\$/ton of N), as well as the product of a nationwide tax rate of \$1/ton of N leached and the leaching intensity (ton of N leached / ton of N applied) that varies by location depending on the ratio of leaching to fertilizer use (Figure S2). After being adjusted by the leaching intensity, the tax imposes a high penalty to the heavy polluters. Their profit margin will be affected directly by the leaching tax and indirectly by the adverse yield impacts of less N fertilizer use. Unlike strategy A that reduces N use and thereby leaching via higher input costs, strategy B achieves the same goal with nationwide implementation of in-field best management practices (BMPs such as precise N application) in order to increase the efficiency of N fertilizer by 10\%. Strategy C and D instead focus on locally feasible BMPs - controlled drainage and wetland restoration. Both practices yield spatially varying leaching removal rates that are determined by local conditions such as water runoff, tile drained area as well as soil and vegetation characteristics etc. These strategies do not affect N fertilizer use directly but remove the pollutant before it enters a stream. Methods to derive the seasonally and spatially varying N removal by controlled drainage and wetland for tile-drained areas are documented in the supplementary materials. 

\section*{Acknowledgement}
We acknowledge support from the USDA-AFRI, grants 2016-67007-24957 and 2019-67023-29679, an NSF-CBET INFEWS/T2 grant 1855937, and an NSF INFEWS/T1 grant 1855996.

\clearpage
\appendix
\section*{Supplementary Materials}
\subsection*{A: Agro-IBIS model}
Agro-IBIS is a comprehensive model of land surface and ecosystem processes that simulates Midwest US natural vegetation (forests, grasslands) and corn, soybean, and wheat agroecosystems, including terrestrial C and N cycling. Crop yields and N loss from the field are determined by agricultural management, environmental stresses on crop development and water balance. Agro-IBIS simulations for corn, soybean, and winter wheat were performed across a 5min by 5min (approximately 8km by 8km) spatial grid across the CONUS. A land mask was created to only simulate grid cells that had more than 1\% cropland area within them based on datasets produced by Ramankutty et al. \cite{ramankutty_farming_2008} and available for download from \url{http://www.earthstat.org/}. This land mask led to 87,590 grid cells simulated by Agro-IBIS.

All scenarios are a “restart run” of 60 years in length using a transient time series of climate and weather for the 1948-2007 time period based on the a daily gridded (5min by 5min) climate dataset \cite{motew2013climate}. Simulations build upon a spin up simulation that spanned 357 years, which was necessary to bring soil biogeochemistry (e.g., coupled C and N cycling) to an equilibrium state. From 1650-1849 natural vegetation was simulated in every grid cell. From 1850-1924, unfertilized continuous winter wheat was simulated across all grid cells. From 1925-2007 continuous corn was simulated, with fertilizer inputs consistent with the previous Donner and Kucharik \cite{donner_corn-based_2008} datasets from 1945-1989, and a switch to \url{http://www.earthstat.org/} fertilizer and manure 5min gridded data for N commencing in 1990. During spin up years from 1650-1947, the 60-year daily climate dataset was “recycled” so that the year 1650 was actual weather year 1948, sequentially stepping through all years in sequence, and then restarting from the beginning when reaching the end of the time series.

While the last 60 years of the restart runs were 2008-2067 in model years, they actually denote climate years of 1948-2007. The first 10 years of the output data is discarded due to a new equilibrium needing to be reached after restart runs commence. Thus, the last 50 years of each simulation can be used for assessment purposes and denote the actual timeseries of climate/weather from 1958-2007. NASA GISS estimates of changing atmospheric $CO_{2}$ from 1650-2007 were used to parameterize Agro-IBIS simulations. During the simulations that approximated contemporary conditions, atmospheric $CO_{2}$ concentration was held fixed at 391 ppm.  

The following logic was used when applying N to cropping systems:
\begin{itemize}

    \item For continuous corn, N from mineral fertilizer and manure was applied at the planting date.
    \item For corn/soybean rotations, N connected to soybeans in the EarthStat.org datasets were applied after fall harvest of soybeans, with N for corn applied at springtime planting date.
    \item For continuous winter wheat, 33\% of N was applied at time of fall planting, with 67\% of N applied on the following January 1. 
    \item For scenarios that represented varied N applications with optimum irrigation practices, a threshold of 50\% plant available water [(actual VWC – PWP VWC)/(FC VWC – PWP VWC)] in the top 60cm of soil was used as the trigger in every grid cell. VWC denotes “volumetric water content”; PWP = permanent wilting point; FC = Field capacity. Irrigation was applied during a nominal 6-hour event to a grid cell and assumed that a maximum daily amount applied is 50mm. Irrigation was only applied to corn phase of rotations and was only turned on when corn was actively growing.

\end{itemize}

The following suite of simulations were performed with Agro-IBIS:
\begin{enumerate}
    \item Rainfed, continuous corn with varied levels of N fertilizer and manure applied (0, 20\%, 60\%, 80\%, 90\%, 100\%, 110\%, 120\%, 140\%, and 180\% of optimum)
    \item Continuous corn with optimum irrigation (50\% plant available water threshold) and varied levels of N fertilizer and manure applied (0, 20\%, 60\%, 80\%, 90\%, 100\%, 110\%, 120\%, 140\%, and 180\% of optimum).
    \item Rainfed corn/soybean rotation, with varied levels of N fertilizer and manure applied to corn (0, 20\%, 60\%, 80\%, 90\%, 100\%, 110\%, 120\%, 140\%, and 180\% of optimum).
    \item Corn/soybean rotation with optimum irrigation (50\% plant available water threshold) and with varied levels of N fertilizer and manure applied to corn (0, 20\%, 60\%, 80\%, 90\%, 100\%, 110\%, 120\%, 140\%, and 180\% of optimum).
    \item Rainfed continuous winter wheat with varied levels of N fertilizer and manure applied (0, 20\%, 60\%, 80\%, 90\%, 100\%, 110\%, 120\%, 140\%, and 180\% of optimum).
    \item Continuous corn with optimum fertilizer across a continuum of irrigation threshold changes (0.2, 0.3, 0.4, 0.5, 0.6, 0.7, and 0.8 available water content to trigger an irrigation event).
    \item Corn/soybean rotation with optimum fertilizer across a continuum of irrigation threshold changes (0.2, 0.3, 0.4, 0.5, 0.6, 0.7, and 0.8 available water content to trigger an irrigation event).
\end{enumerate}

\subsection*{B: SIMPLE-G-US-CS model}
\subsubsection*{Main structure of the model}  SIMPLE-G-US-CS is a global partial equilibrium economic model focusing on grid-specific crop productions and agricultural inputs use. It is a special version of the SIMPLE-G-US model \cite{baldos2020simple} but includes two crops - corn and soybeans. At each grid cell, crop production follows a multi-nesting output structure shown in Figure \ref{fig:fs-1}. Each nesting represents a production function with the constant elasticity of substitution (CES) form. The nesting structure differs at the bottom layer between irrigated and rainfed production. For irrigated crops, irrigation water is combined with irrigable land to produce a land-water composite, which is further combined with non-land inputs (e.g. capital and labor) to produce an augmented land input that is finally combined with N fertilizer to produce the ultimate crop output. The N fertilizer application rate data was provided by Lark et al. \cite{TylerLark2021}.

The parameters designated by the sigmas are the elasticity of substitution between the two inputs within each layer. A larger value of sigma indicates easier substitution between inputs. When the use of one input is restricted by policy or natural constraints, the other input from the same layer will be used more if the elasticity is large. As a result,  the impact on output will be smaller because of the more flexible production technology. The production function including cost shares and substitutability of inputs is unique for each grid-cell.

The demand for crop commodities is driven by biofuels production, population, and per capita income. The equilibrium for sluggish and semi-mobile inputs including cropland and water is established at the grid-cell level; while the markets for mobile inputs such as labor and capital are cleared at the regional level. More descriptions of the SIMPLE-G-US model can be found in Baldos et al. \cite{baldos2020simple} and Sun et al. \cite{sun2020fine}.

\begin{figure}
\setcounter{figure}{0}
\makeatletter 
\renewcommand{\thefigure}{S\@arabic\c@figure}
\makeatother

\centering
\includegraphics[width=\textwidth]{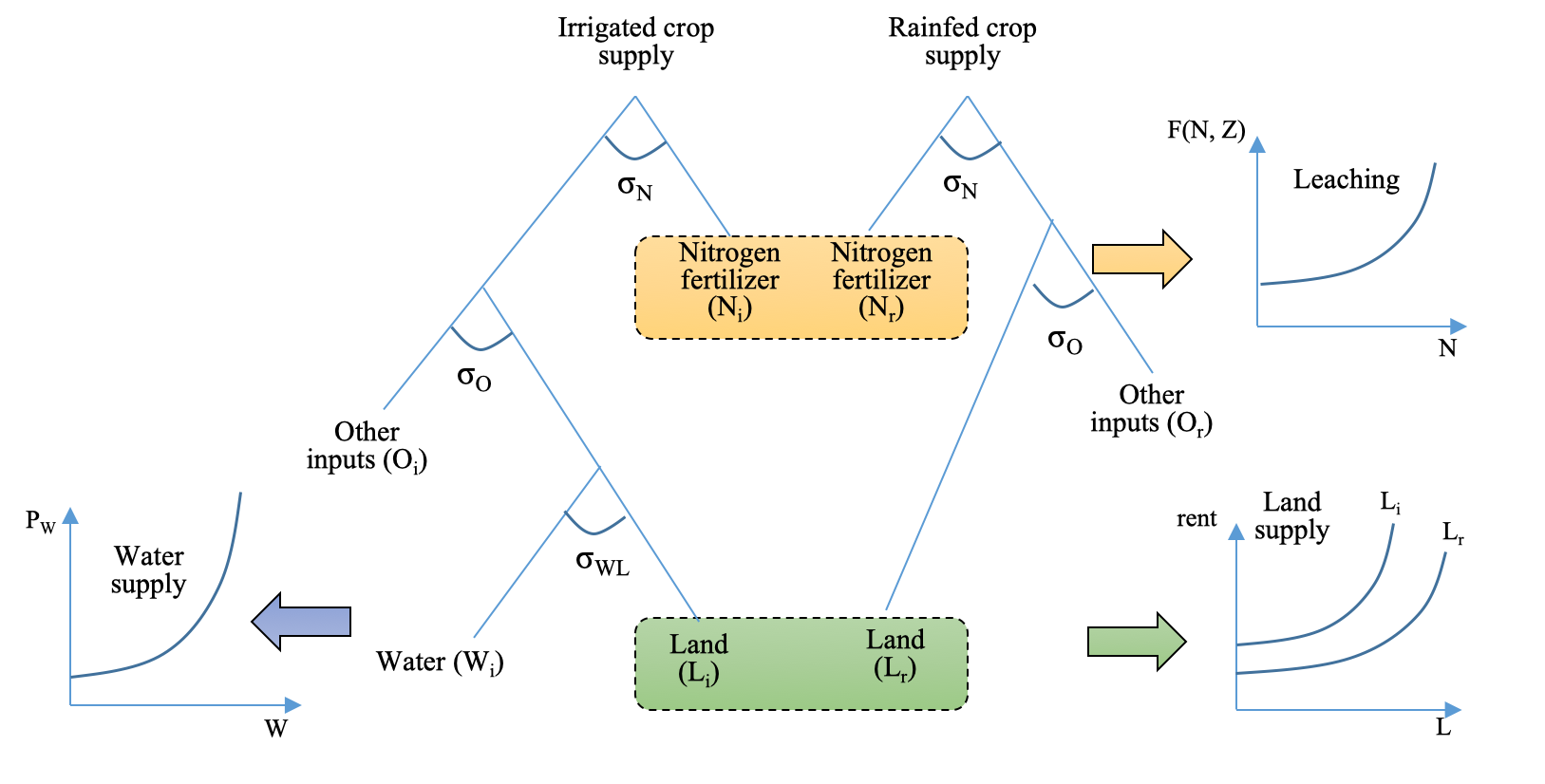}
\caption{A schematic of the supply side of SIMPLE-G-US-CS model.}
\label{fig:fs-1}
\end{figure}

\subsubsection*{Nitrate leaching module} 
\begin{figure}
\centering
\makeatletter 
\renewcommand{\thefigure}{S\@arabic\c@figure}
\makeatother

\includegraphics[width=\textwidth]{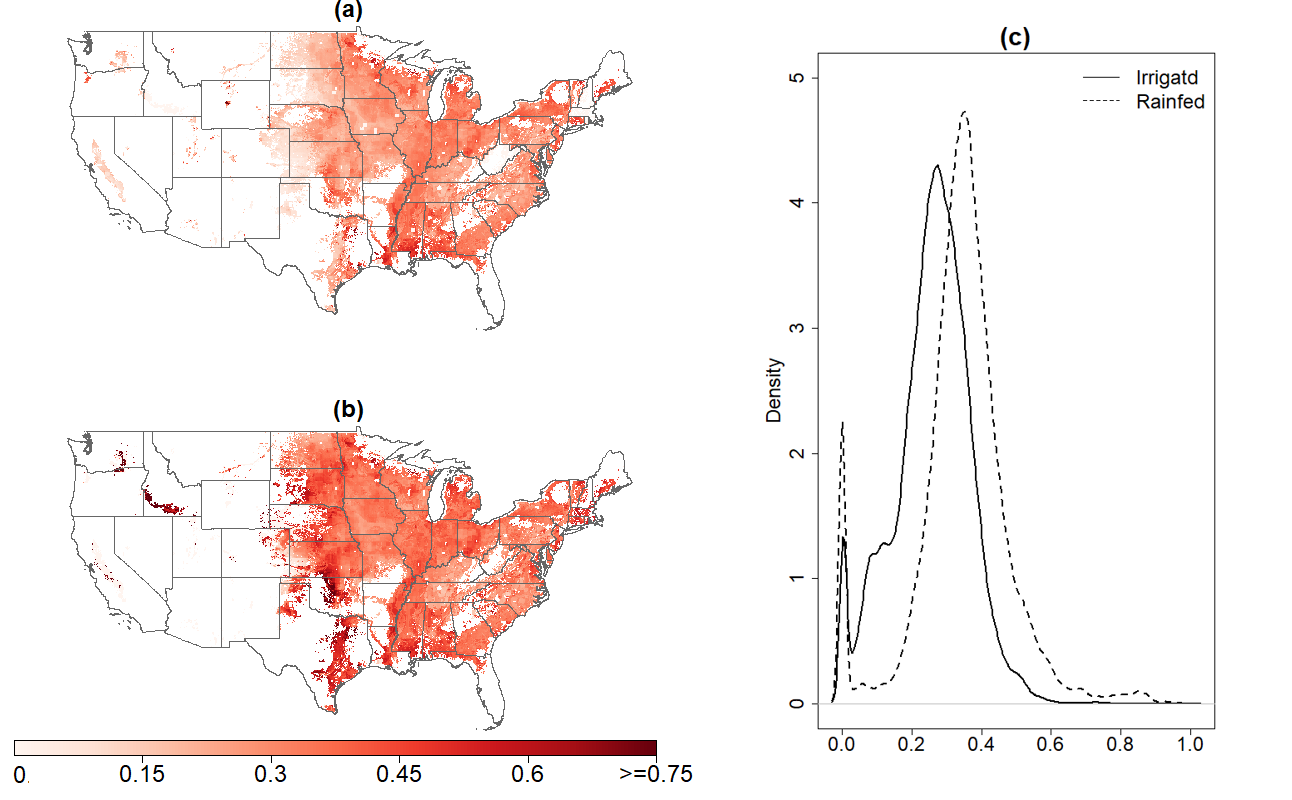}
\caption{N leaching intensity (between 0 to 1) for (a) irrigated and (b) rainfed crop production, computed based on the ratio of leaching rate (kg/ha) to N fertilizer application rate (kg/ha) \cite{TylerLark2021}. Figure S2-(c) in the panel shows the density plot of the N leaching intensity for irrigated and rainfed production.}
\label{fig:fs-2}
\end{figure}

In order to link nutrient management practices and policies to crop production, a N leaching module is added to SIMPLE-G-US-CS. The grid-specific leaching response (kg of N leaching per ha) to N fertilizer application rate (kg of N applied per ha) follows a quadratic functional form to reflect the non-linear relationship between nutrient application and leaching. The coefficients of the quadratic function are estimated by fitting a transfer function through the N application and leaching outcomes simulated by the Agro-IBIS model. The ratio of the leaching response (kg/ha) to N fertilizer application rate (kg/ha) ranges between zero and one. We define this ratio as `leaching intensity'. A higher leaching intensity indicates a larger proportion of applied N fertilizer that is lost in the stream. Figure \ref{fig:fs-2} visualizes the leaching intensity (ranges between 0 and 1) for irrigated and rainfed corn production based on the baseline data that we constructed for the leaching module. The density plot shows that both intensities follow some normal distribution. The leaching intensity concentrates around 0.2-0.5, indicating that for most grid-cells 20\% to 50\% of the N fertilizer applied will be lost. On average, the leaching intensify for rainfed practice is higher than that of irrigated. 

This leaching intensity plays a key role in determining the "penalty" for polluting, as explained by Equation \ref{eq:leachingcharge}. In our model, the leaching intensity ($\theta_g$) incurs an additional cost ($\theta_g\times t_r$), which adjusts the uniform leaching tax $t_r$ (in \$/kg of N leaching) such that the grid-cell with a higher leaching intensity (i.e. the "leaky" soil) faces a higher N fertilizer cost. Although both the market price of N fertilizer and the leaching charge are uniform nationwide, the leaching intensity varies by grid-cell, driving the grid-cell level heterogeneity in the actual cost of using N fertilizer. 
\begin{equation}
    P_{N,g} = P_{N,r} + \theta_g\times t_r
    \label{eq:leachingcharge}
\end{equation}
where $r$ indicates country, and $g$ indicates grid-cell.

\begin{figure}[tbhp!]
\centering
\makeatletter 
\renewcommand{\thefigure}{S\@arabic\c@figure}
\makeatother

\includegraphics[width=\linewidth]{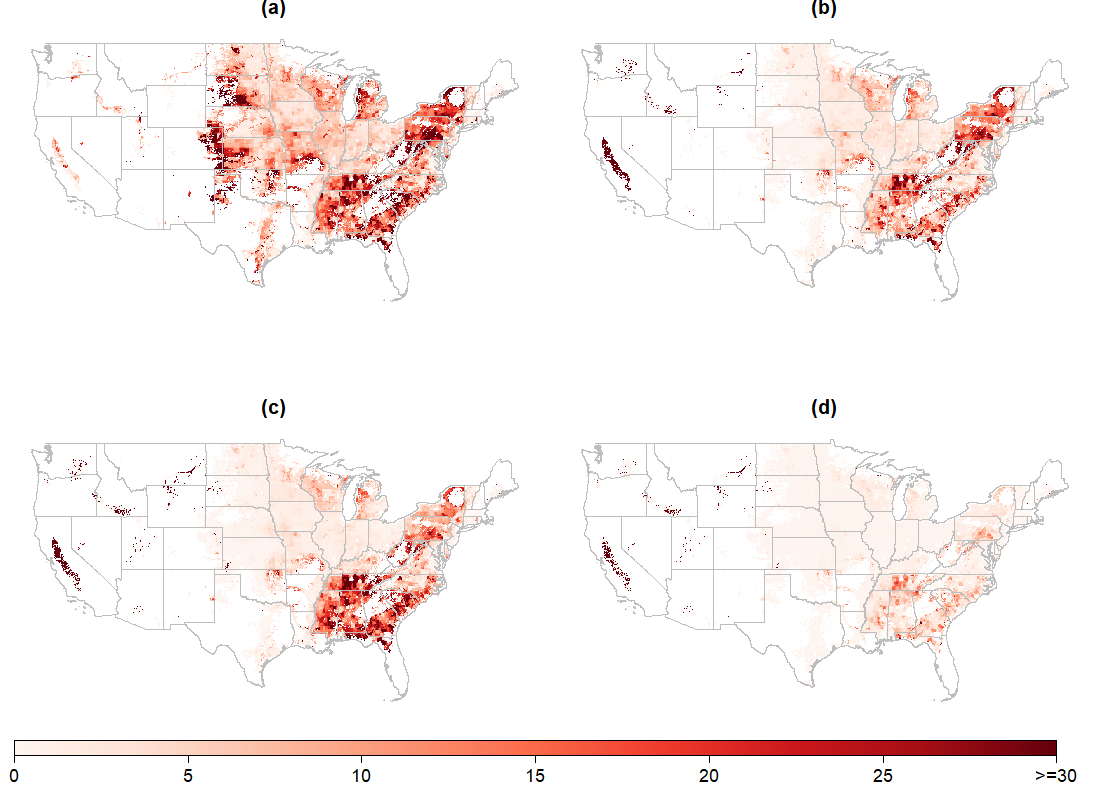}
\caption{Marginal product of N fertilizer measured by the increased output associated with an additional unit of N fertilizer use. Unit is tons of crop output per additional ton of N fertilizer use, evaluated at the grid-specific N fertilizer application rate provided by the baseline data. For example, the grid-cell of a value of 20 means that one more ton of N fertilizer use that costs \$210 (at model baseline price) yields 20 more tons of crop output that values \$535.6 (at model baseline corn price of \$26.78/ton). Maps show the marginal product of N under (a) irrigated continuous corn, (b) rainfed continuous corn, (c) irrigated corn-soy rotation, and (d) rainfed corn-soy rotation.}
\label{fig:fs-3}
\end{figure}

How much the leaching "penalty" could affect farmers' decision of cutting back N fertilizer application depends also on the marginal product of N fertilizer, i.e. the additional output that yields from one more unit of N fertilizer application, as shown in Figure \ref{fig:fs-3}. The marginal product of N fertilizer use determines the optimal amount of N fertilizer to be applied. At a certain location, this marginal product declines as N fertilizer application increases, because of the diminishing marginal productivity of inputs. In the absence of a leaching tax, farmers apply fertilizer to increase crop yields up to an equilibrium point where marginal output revenue equals marginal cost, as long as the amount of N fertilizer application has not reached the N fertilizer breaking point (the application rate beyond which crop yield declines). With leaching tax, the equilibrium is achieved at a lower level of N fertilizer application where the marginal product of N is higher in order to match the higher unit price of N fertilizer application. A larger marginal product of N fertilizer indicates a greater loss of crop output per unit of N fertilizer use reduction, thereby more sensitive yield response to the leaching tax. Figure \ref{fig:fs-3} shows that this marginal product of N fertilizer is typically higher in eastern side of the fringe area than in the western side. That explains why leaching tax appears more effective than nutrient management in the east of the US (Figure 3 in the main text).

\subsubsection*{Model validation}
We validate the model by comparing observed and simulated changes over the period of 1990-2006. The strategy is to target an end year crop price and let the model endogenously determine how much the quantity variables need to change in order to achieve the targeted price. Then we compare the simulated changes with historical data to make sure they agree reasonably well at least at the national level. The focal variables for this validation process include the harvested acreage of corn and soybeans, the quantity of N fertilizer application, and the quantity of corn and soybeans output. The 1990-2006 time horizon is selected based on several considerations. Some data required for this validation exercise are only available from circa 1990 but not earlier. From the modeling point of view, SIMPLE-G-US-CS is more suitable for mid- to long-term than for short-term projections. Finally, we have sought to avoid the years characterized by unusual market movement caused by e.g. extreme weather and policy changes. For example, one important development relevant to N fertilizer use is the soaring price of natural gas in the first decade of the 21st century and in the year of 2021. Because energy prices are not determined by the model, this change in fertilizer price is considered as an external shock to the baseline. A similar argument applies to technological change that increases the total factor productivity of agriculture during this period, which is also treated as an external perturbation to the system. 

\subsection*{C: Agro-IBIS and SIMPLE-G-US-CS coupling} 
\subsubsection*{Transfer functions}
Using the simulated yield outcomes at the various N fertilizer rates provided by the Agro-IBIS model, we fit a Gompertz function to these data points at each grid-cell. These fitted curves, defined as "transfer functions", translate the Agro-IBIS output into a univariate relationship between crop yield ($f_g$, kg/ha) and normalized N fertilizer use ($n$, N application rate in kg/ha), shown in Equation \ref{eq:gompertz}:

\begin{equation}
    f_g(n)=a_g\times e^{-b_gc_g^{n_g}}
    \label{eq:gompertz}
\end{equation}
where $f()$ is the normalized production function, $n$ is the fertilizer application rate $\frac{N}{L}$ in kg/ha. Parameters $a$, $b$, $c$ were estimated by fitting the Gompertz function. Coefficients $a$, $b$, $c$, as well as $f(n)$ and $n$ all vary by grid-cell and therefore index over $g$ (indicating grid-cells). The values are unique for each of the 48,317 US grid-cells in the SIMPLE-G-US-CS model. 
\subsubsection*{Derive the elasticity of substitution from a normalized production function}
According to Ferguson \cite{ferguson2008neoclassical} (page 97, Equation 5.2.18), the elasticity of substitution between N fertilizer and land $\sigma_{LN}$ can be expressed as a function of $F$ (output), $F_L$ and $F_N$ (the first derivative of output with respect to land and N fertilizer input respectively), and $F_{LN}$ (the second derivative of output with respect to land and N fertilizer). 
\begin{align}
\sigma_{LN}&=\frac{F_L\times F_N}{F\times F_{LN}}
\label{eq:sigma}
\end{align}
where $L$ indicates the harvested area of crops in hectares and $N$ indicates N fertilizer application in kg. All elements in the equation are grid-cell specific. We omit the grid-cell notation `g' for simplicity. In order to utilize the crop yields (kg/ha) and Nitrogen fertilizer use (kg/ha) provided by Agro-IBIS model, total output is broken into cropland area $L$ and yields $f(n)$, where $n$ is the N fertilizer application rate $N/L$ in kg/ha and $f(n)$ is the normalized production function shown in Equation \ref{eq:gompertz}. 
\begin{equation}
F=L \times f(\frac{N}{L})=L \times f(n)
\label{eq:gompertz}
\end{equation}

\begin{figure}[tbhp]
\centering
\makeatletter 
\renewcommand{\thefigure}{S\@arabic\c@figure}
\makeatother

\includegraphics[width=\linewidth]{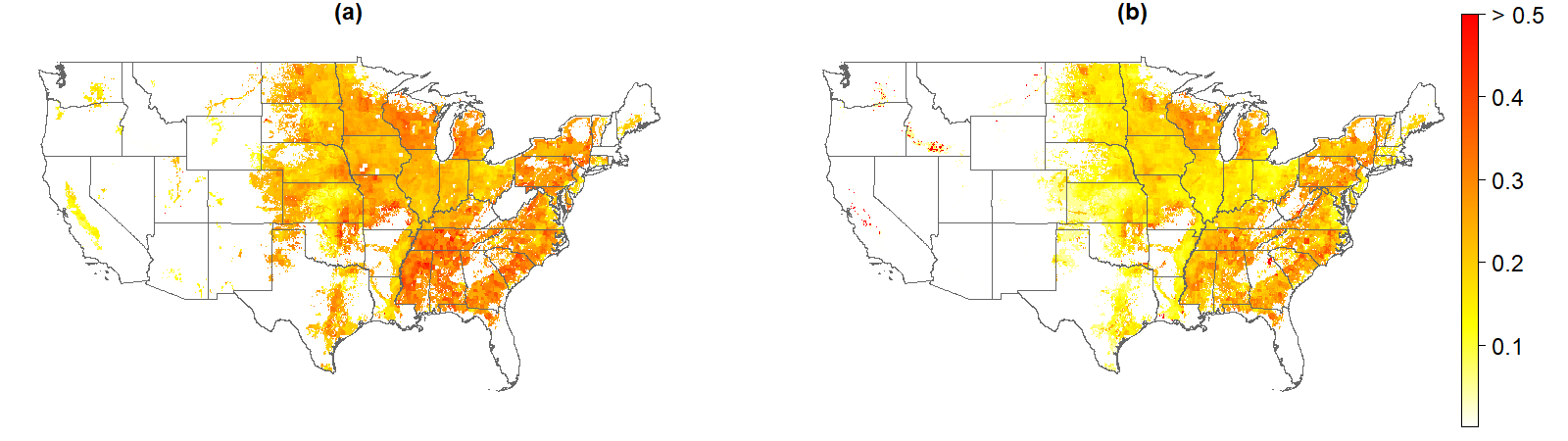}
\caption{Elasticity of substitution between N fertilizer and augmented land input for (a) irrigated and (b) rainfed crop production. }
\label{fig:fs-4}
\end{figure}

The elasticity of substitution $\sigma_{LN}$ in Equation \ref{eq:sigma} is then transformed to a function of $f$, $f'(n)$, $f''(n)$ and $n$.

\begin{align}
\sigma_{LN }&=\frac{F_L\times F_N}{F\times F_{LN}}\\
			&=\frac{[f(n)-\frac{N}{L}f'(n)]\times f'(n)}{L\times f(n) \times (-\frac{N}{L^2})f''(n)}\\
            &=\frac{f'(n)[f(n)-nf'(n)]}{-nf''(n)f(n)}
            \label{eq:sigma-uni}
\end{align}

\begin{equation}
F_L=f(n)+L\times f'(n)\times(-\frac{N}{L^2})=f(n)-\frac{N}{L}f'(n)
\end{equation}
\begin{equation}
F_N=L\times f'(n)\times \frac{1}{L}=f'(n)
\end{equation}
\begin{equation}
F_{LN}=F_{NL}=f'(n)\times\frac{1}{L}-\frac{N}{L}f''(n)\times\frac{1}{L}-\frac{1}{L}f'(n)=-\frac{N}{L^2}f''(n)
\end{equation}

\begin{align}
&f(n)=a\times e^{-bc^n} \label{d1}\\
&f'(n)=a\times e^{-bc^n}(-bc^n)\ln(c) \label{d2}\\
&f''(n)=a\times e^{-bc^n}(-bc^n)\ln^2(c)(-bc^n+1) \label{d3}
\end{align}
Given values for $a$, $b$, $c$, and $n$, one can evaluate $f$, $f'(n)$, and $f''(n)$ based on Equation \ref{d1}, \ref{d2}, and \ref{d3}. Plugging $f$, $f'(n)$, and $f''(n)$ into Equation \ref{eq:sigma} yields the elasticity of substitution $\sigma_{LN}$ between N fertilizer and augmented land inputs by grid-cell, as shown in Figure \ref{fig:fs-4}. The elasticity for irrigated production is generally larger than that of rainfed production. Also the substitubability between N and land inputs is lower in the heart of the Corn Belt. 

\subsection*{D: Seasonally and spatially-variable Nitrate leaching function for tile-drained areas}

Strategy C (controlled drainage) and D (wetland restoration) both require spatially varying constraints within the SIMPLE-G-US-CS framework to represent differences in site suitability and practice effectiveness. This section explains our methods to determine the feasibility and the effectiveness of N leaching mitigation of controlled drainage and functional wetland restoration.

\subsubsection*{Site Suitability} 
The data sources used to determine site suitability for controlled drainage and wetland restoration were the 2017 gSSURGO data from the Natural Resources Conservation Service \cite{mittergridded}, hydric soil data layer, and the 2015 Cropland Data Layer from the National Agricultural Statistics Service \cite{USDANASS}.

\subparagraph*{Controlled Drainage} 
The potential area for implementation of controlled drainage (CD) for the Upper Midwestern states (Illinois, Iowa, Indiana, Minnesota, Missouri, North Dakota, Ohio, South Dakota and Wisconsin) was estimated by Bob Dobos, Soil Scientist at the National Soil Survey Center. This product identifies lands that are likely to be drained for crop production, on which controlled drainage would be economically feasible. The 30 m resolution map identifies areas with: flat topography (slope 1\% or less), soils that have a high-water table during the growing season (saturated to 46 cm or less during the growing season using hydric soil criteria), cropland and 15 acres or more of contiguous surface area (to represent economic feasibility). The 30 m categorical dataset (suitable/not suitable for CD) was unprojected to a geographic coordinate system and aggregated to find the fraction of each 5 minute grid cell that was suitable for CD.

\subparagraph*{Wetland Restoration}
The potential area for cropland that could be taken out of production and converted back to a wetland land use was limited to land areas within the Mississippi River basin that are currently farmed with hydric soils. The 30 m categorical dataset (suitable/not suitable for wetland restoration) was unprojected to a geographic coordinate system and aggregated to find the fraction of each 5 arc-minute grid cell that was suitable for wetland restoration.

\subsubsection*{Nitrate Loss Reduction} 
The Agro-IBIS simulations of corn and soybeans provide two spatially-varying functions to SIMPLE-G-US-CS describing N leached (kg/ha/year) vs. kg applied nitrogen and the corn equivalent yield (kg/ha/year) vs. kg applied nitrogen. The N leached as predicted by Agro-IBIS can be considered the nitrate load without controlled drainage (Scenario C) or into the wetland (Scenario D), and here we define spatially-varying adjustments expressed as a percent of the unmodified load ($100\times L_{adjusted}/L_{original}$).

\subparagraph*{Controlled Drainage} 
Helmers et al. \cite{Helmers2021} synthesized field results from 14 sites in Ohio, Indiana, Illinois, Iowa and Minnesota to evaluate the impact of controlled drainage on nitrate load, by season and by USDA Plant Hardiness Zone. They found that there was no reduction in nitrate concentration with Controlled Drainage (CD), so any reduction in Nitrate load was due to a reduction in drainflow relative to the Freely Drained (FD) condition. The change in annual drainflow (CD-FD, measured in mm/year) was not statistically different between Plant Hardiness Zones.  However, the percent reduction in drainflow does vary by seasons, and since drainflow seasonality also varies by location this can result in some spatial variation in estimated annual reductions.  Therefore, the CD volume was calculated as a function of monthly simulated FD volume, with different CD ratios for each season, as shown in Table \ref{tab:cdr}. Note that the values presented in Table \ref{tab:cdr} were calculated based on the average drainflow reduction by season across all sites analyzed by Helmers et al. \cite{Helmers2021}, but the final paper reported median values, not means.

\begin{table}[!ht]
\centering
\caption{Seasonal Controlled Drainage Ratio (CDR=CD/FD)}
\begin{tabular}{ccccc}
\hline
 & Jan-Mar & Apr-Jun & Jul-Sep & Oct-Dec  \\
 & 36.4 & 57.2 & 54.8 & 81.7 \\
 \hline
 \label{tab:cdr}
\end{tabular}
\end{table}

\begin{figure}[tbhp]
\centering
\makeatletter 
\renewcommand{\thefigure}{S\@arabic\c@figure}
\makeatother

\includegraphics[width=.55\linewidth]{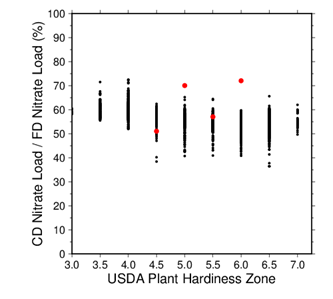}
\caption{Estimated CD Nitrate load as a percent of FD nitrate load along with aggregated observations from Helmers et al. \cite{Helmers2021}, in red.}
\label{fig:fs-5}
\end{figure}

\begin{figure}[tbhp]
\centering
\makeatletter 
\renewcommand{\thefigure}{S\@arabic\c@figure}
\makeatother

\includegraphics[width=.55\linewidth]{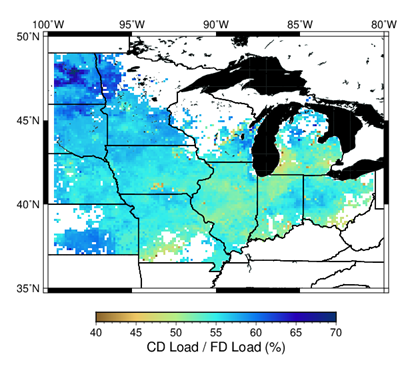}
\caption{Estimated CD Nitrate load as a percent of FD nitrate load from drained agricultural land based on a seasonal percent reduction in drainflow.}
\label{fig:fs-6}
\end{figure}

The spatially-varying annual adjusted load as a percent of the original load from Agro-IBIS is therefore calculated as:
\begin{equation}
    LoadAdjustment = 100\% \times \frac{\sum_{i=1}^{12}Q_{cl,i}\times CDR_i}{\sum_{i=1}^{12}Q_{cl,i}}
    \label{eq:loadadj}
\end{equation}
Where $Q_{cl,i}$ is the drainage rate from corn and soy for month i (m/day) supplied from monthly average depths of subsurface drainage simulated by the Variable Infiltration Capacity (VIC) model \cite{lee2017regional} and $CDR_i$ is the seasonal CD as a percent of FD from Table \ref{tab:cdr}. The distribution of annual CD Nitrate load as a percent of FD load estimated by Equation \ref{eq:loadadj} is shown in Figure \ref{fig:fs-5}, along with observed reductions from Helmers et al. \cite{Helmers2021}.  Figure \ref{fig:fs-6} shows the resulting spatial distribution of CD ratio values that were input to the SIMPLE-G-US-CS model for Scenario C.

\subparagraph*{Wetland Restoration} 
A relationship describing Nitrate load out of the wetland (kg/ha, $L_{out}$,) as a function of wetland area was derived based on a tanks-in-series model.  This analysis assumes that all water leached from the root zone could be captured by a wetland.  
Hydraulic loading rate to a wetland is defined as:

\begin{equation}
    hlr_i=\frac{q_{d,i}}{A_w}\times Q_{cl,i}=\Bigg(\frac{1-f_w}{f_w}\Bigg)\times Q_{cl,i}
\end{equation}

Where, $hlr$ = Hydraulic loading rate; $q_{d,i}$ = daily wetland inflow for month i (in m3/day) which is $A_{cl}\times Q_{cl,i}$. $Q_{cl,i}$ is the drainage rate from corn and soy for month i (m/day) supplied from monthly average depths of subsurface drainage simulated by the VIC model \cite{lee2017regional}. $A_{cl}$ is the area of subsurface drained corn and soy contributing to the wetland and $A_w$ is the area of wetland (both in $m^2$). The ratio of wetland area required to treat a given cropland area $f_w=\frac{A_w}{A_{cl}}$, and a value of $f_w$=0.005 was assumed for this work.

The maximum drainage extent, $A_d$, was determined for each grid cell based on the fraction of the land area in corn or soybean production based on the 2015 Cropland Data Layer (CDL) and a soil drainage class of somewhat poorly drained, poorly drained or very poorly drained.  

Crumpton \cite{crumpton2001using} applied a mass balance nutrient model to evaluate the impact of agricultural treatment wetlands across the Midwest that utilized a tanks-in-series model, as follows:

\begin{equation}
   \frac{C_o-C_b}{C_i-C_b}=\Bigg(1+\frac{K_i}{N\times hlr}\Bigg)^{-N}
\end{equation}

Where $C_i$ is the inflow concentration, $C_o$ is the outflow concentration and $C_b$ (g/m3) is the background concentration and $K_i$ (m/year) is an areal removal rate constant for month i. As we are interested in nitrate-nitrogen, $C_b$ is 0 \cite{Ward2016}. N is the number of tanks in series, which is generally determined by calibration. For higher values of N, the results of the tanks-in-series model collapses to that of the exponential decay model, we utilized a value of N=1.

$K_i$ will depend on the climate, soil, and vegetation characteristics of each individual wetland.  For this analysis, spatial variability due to seasonal temperature impacts on denitrification rates was modeled as follows Tchobanoglous and Schroeder \cite{tchobanoglous1985water}:

\begin{equation}
   K_i=k_{Deni}\times\theta^{T_i-20}
\end{equation}

Where $k_{Deni}$ is the volumetric rate constant for denitrification (m/d), $\theta$ is a dimensionless temperature rate coefficient, and $T_i$ is the water temperature in °C for month i.  For shallow wetland environments, we will assume that the water temperature is equal to the air temperature, or 0 °C, whichever is larger. The theta parameter ($\theta$) has a range in values from 1.0-1.09 \cite{kadlec2009treatment, mitsch_wetland_2009, crumpton2001using}. $k_{Deni}$ varies from about 0.09 m/day to 0.15 m/day \cite{Ward2016, crumpton2001using}. Value of $\theta$=1.09 and $k_{Deni}$=0.15 m/day were used, based on comparisons with observations, as shown below.

As a conservative estimate of the potential load reduction due to wetlands, we assume that water losses in the treatment wetland are negligible. The flow rate out of the wetland, on average, is equal to the flow rate into the wetland, and the load equation is equivalent to the concentration equation: 

\begin{equation}
   Lo_i=Li_i\times\Bigg(1+\frac{K_i\times f_w}{N\times(1-f_w)\times Q_{cl,i}}\Bigg)^{-N}
\end{equation}
Where $L_i$ is the leaching function from Agro-IBIS (in kg/ha/year).  Assuming that the inflow concentration is constant over all the months, the spatially-varying annual adjusted load as a percent of the original load from Agro-IBIS for the tanks-in-series model is therefore calculated as:
\begin{equation}
   LoadAdjustment = 100\% \times \frac{\sum_{i=1}^{12}Lo_{i}}{\sum_{i=1}^{12}Li_{i}} = 100\% \times \frac{\sum_{i=1}^{12}\Bigg[Q_{cl,i}\times\Bigg(1+\frac{K_i\times f_w}{N\times(1-f_w)\times Q_{cl,i}}\Bigg)^{-N}\Bigg]}{\sum_{i=1}^{12}Q_{cl,i}}
\end{equation}

Crumpton et al. \cite{Crumpton2006} identified that HLR was a key variable in explaining the difference in Nitrate removal rates for wetlands observed across Ohio, Illinois and Iowa, based on data from several other researchers \cite{davis1981prairie, kovacic2000effectiveness, phipps1994factors, phipps1997nitrate, mitsch2005nitrate,zhang2000hydrologic,zhang2000water,Zhang2002,zhang2004water}.  Figure \ref{fig:fs-5} shows the observed outlet Nitrate load as a percent of influent load from the synthesis by Crumpton et al. \cite{Crumpton2006}, as well as the simulated values for this study using the mean monthly cycle (1981-2010) of drainflow depth from Lee \cite{lee2017regional} and observed monthly average air temperature prepared by \cite{maurer2002long}, both at 12.5 arc-minute resolution. Figure \ref{fig:fs-6} shows the percent reduction values that were input to the SIMPLE-G-US-CS model for Scenario D.

\begin{figure}
\centering
\makeatletter 
\renewcommand{\thefigure}{S\@arabic\c@figure}
\makeatother

\includegraphics[width=.55\linewidth]{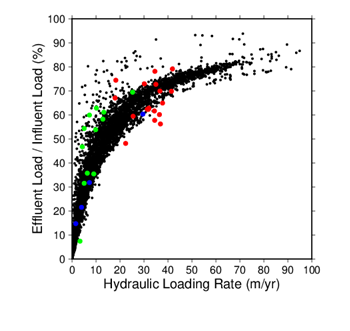}
\caption{Comparison of the temperature-dependent Tanks-in-Series Model implemented using monthly average drainflow and air temperature with N=1, $\theta$=1.09, $K_d$=0.15 m/day and $f_w$=0.005 (black dots) to observations from Ohio (red), Illinois (green) and Iowa (blue) based on data summarized by Crumpton et al. \cite{Crumpton2006}.}
\label{fig:fs-7}
\end{figure}

\begin{figure}
\centering
\makeatletter 
\renewcommand{\thefigure}{S\@arabic\c@figure}
\makeatother

\includegraphics[width=.55\linewidth]{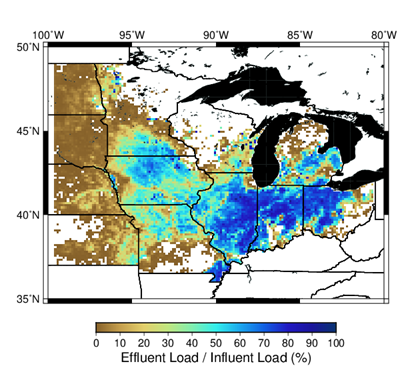}
\caption{Estimated effluent Nitrate load as a percent of wetland influent nitrate load from drained agricultural land using the Tanks-in-Series Model with N=1, $\theta$=1.09, $K_d$=0.15 m/day and $f_w$=0.005}
\label{fig:fs-8}
\end{figure}

\begin{figure}
\centering
\makeatletter 
\renewcommand{\thefigure}{S\@arabic\c@figure}
\makeatother

\includegraphics[width=.78\linewidth]{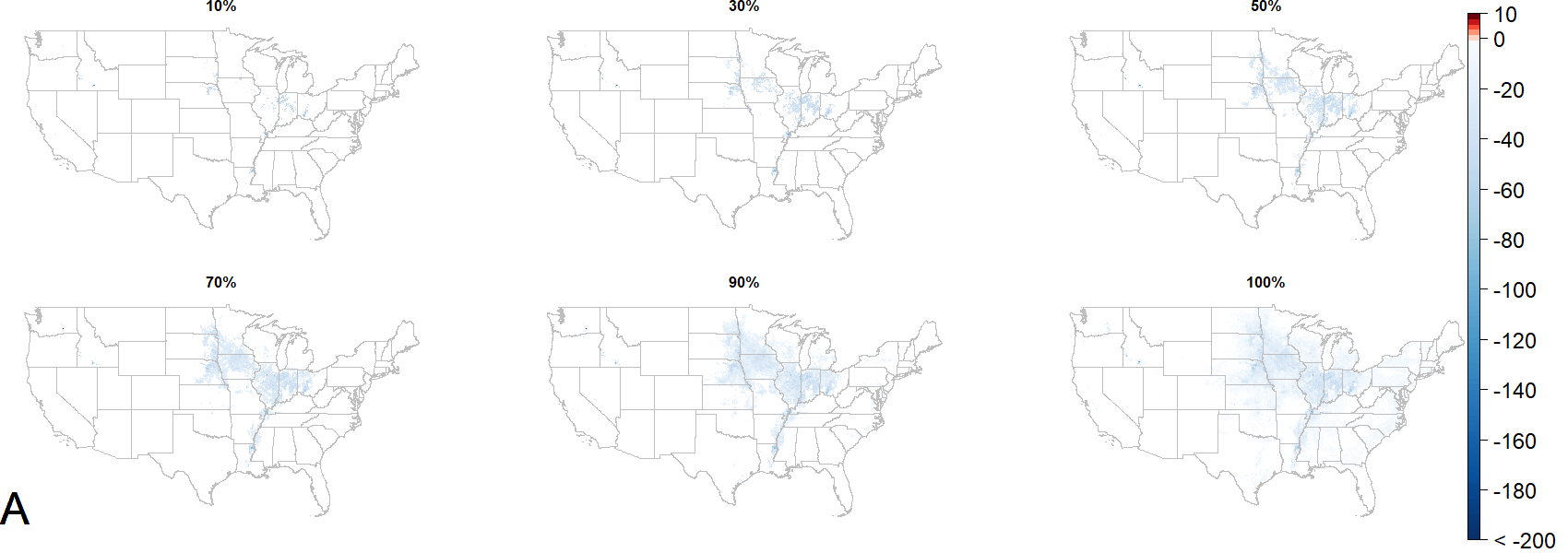}
\vspace{1cm}

\includegraphics[width=.78\linewidth]{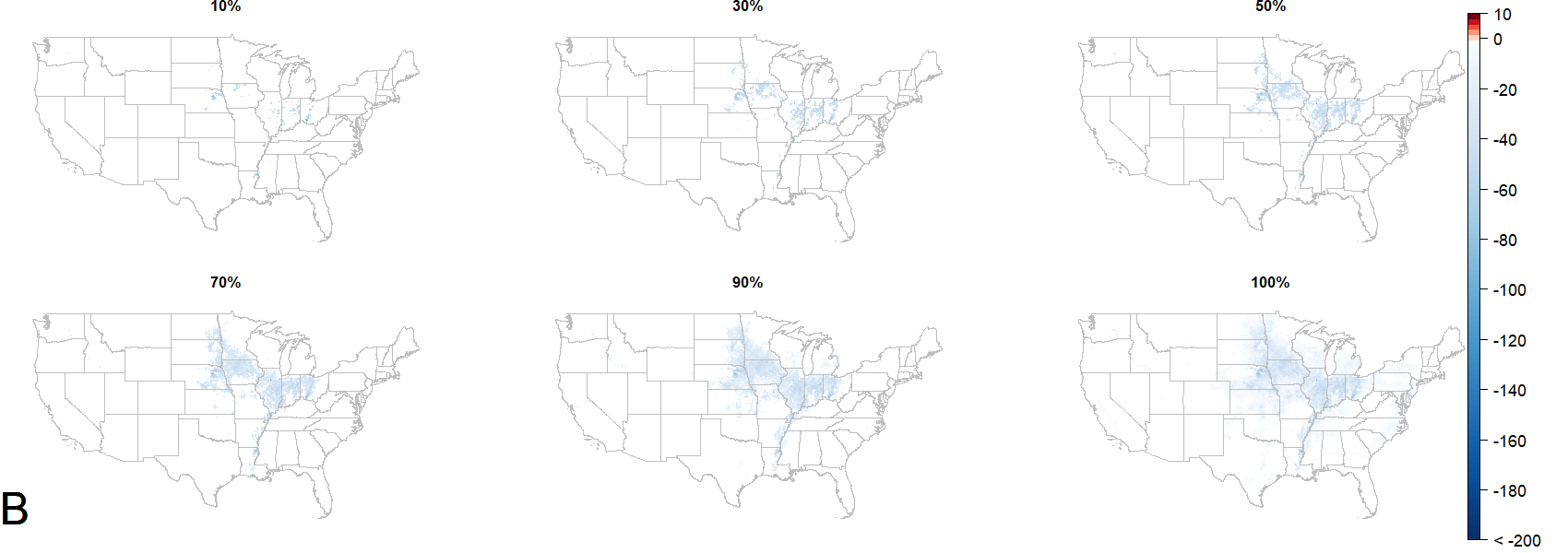}
\vspace{1cm}

\includegraphics[width=.78\linewidth]{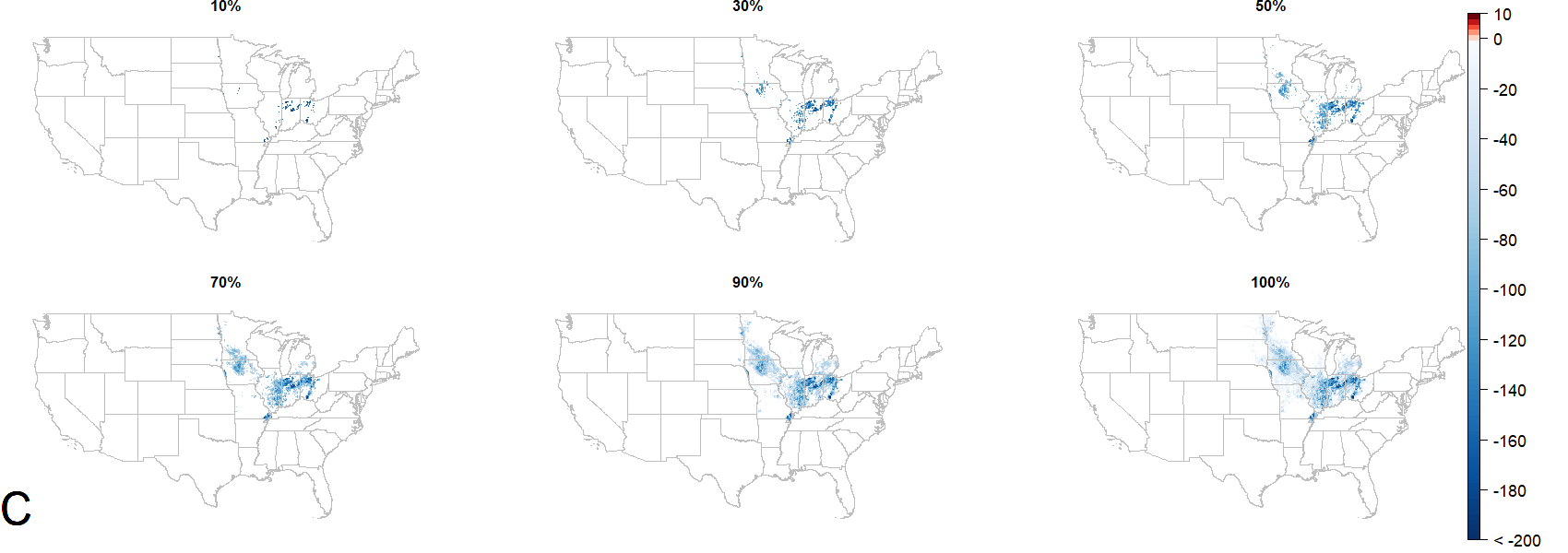}
\vspace{1cm}

\includegraphics[width=.78\linewidth]{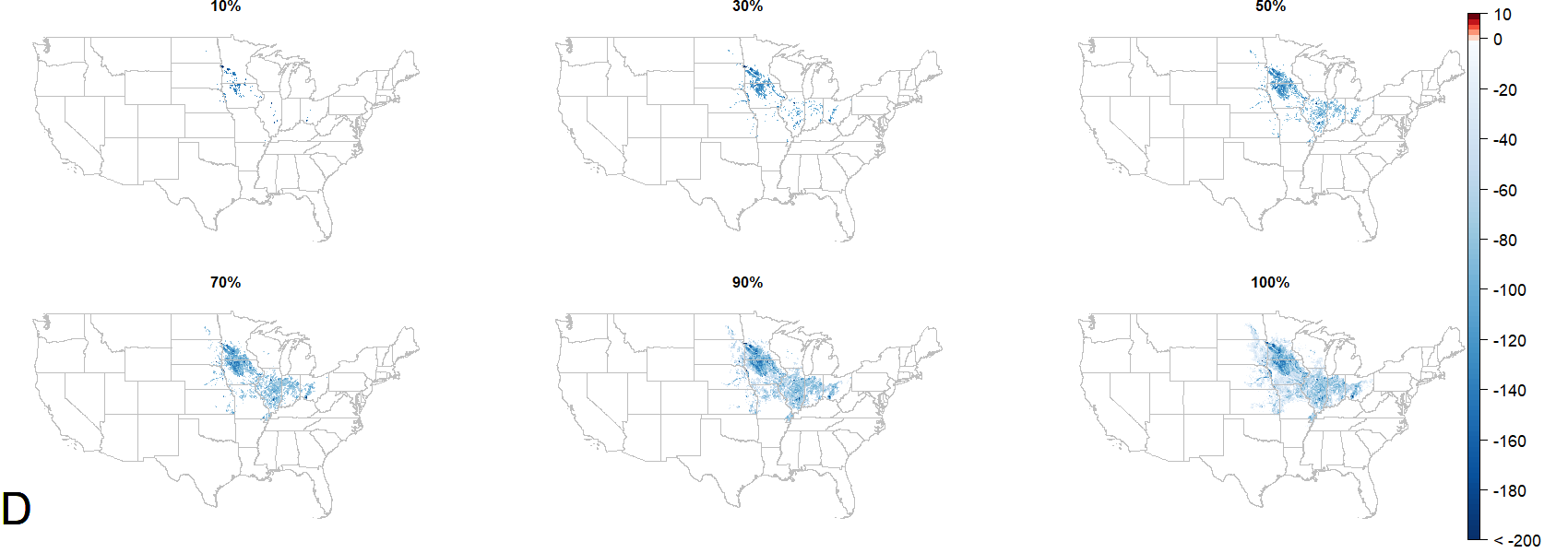}
\caption{Locations that contribute 10\%, 30\%, 50\%, 70\%, 90\%, 100\% of total mitigation under strategies of A, B, C and D. Colors indicate the changes of N leaching as results of the N mitigation policies. Unit is tons of N leaching per grid-cell. Negative values indicates N leaching reduction.}
\label{fig:fs-9}
\end{figure}

\clearpage
\bibliography{pnas-ref.bib}

\begin{thebibliography}{10}

\bibitem{turner_emergence_2007-1}
B.~L. Turner, Eric~F. Lambin, and Anette Reenberg.
\newblock The emergence of land change science for global environmental change
  and sustainability.
\newblock {\em Proceedings of the National Academy of Sciences},
  104(52):20666--20671, December 2007.

\bibitem{goolsby_nitrogen_2001}
D.~A. Goolsby, W.~A. Battaglin, B.~T. Aulenbach, and R.~P. Hooper.
\newblock Nitrogen input to the {Gulf} of {Mexico}.
\newblock {\em Journal of Environmental Quality}, 30(2):329--336, March 2001.

\bibitem{vitousek_human_1997}
Peter~M. Vitousek, John~D. Aber, Robert~W. Howarth, Gene~E. Likens, Pamela~A.
  Matson, David~W. Schindler, William~H. Schlesinger, and David~G. Tilman.
\newblock Human {Alteration} of the {Global} {Nitrogen} {Cycle}: {Sources} and
  {Consequences}.
\newblock {\em Ecological Applications}, 7(3):737--750, 1997.

\bibitem{diaz_spreading_2008}
Robert~J. Diaz and Rutger Rosenberg.
\newblock Spreading {Dead} {Zones} and {Consequences} for {Marine}
  {Ecosystems}.
\newblock {\em Science}, 321(5891):926--929, August 2008.

\bibitem{rabalais_hypoxia_2001}
N.~N. Rabalais, R.~E. Turner, and W.~J. Wiseman.
\newblock Hypoxia in the {Gulf} of {Mexico}.
\newblock {\em Journal of Environmental Quality}, 30(2):320--329, March 2001.

\bibitem{USEPA2013}
{\em Reassessment 2013: Assessing Progress Made Since 2008}, 2013.

\bibitem{scavia2017ensemble}
Donald Scavia, Isabella Bertani, Daniel~R Obenour, R~Eugene Turner, David~R
  Forrest, and Alexey Katin.
\newblock Ensemble modeling informs hypoxia management in the northern gulf of
  mexico.
\newblock {\em Proceedings of the National Academy of Sciences},
  114(33):8823--8828, 2017.

\bibitem{mclellan_right_2018}
Eileen~L. McLellan, Keith~E. Schilling, Calvin~F. Wolter, Mark~D. Tomer,
  Sarah~A. Porter, Joe~A. Magner, Douglas~R. Smith, and Linda~S. Prokopy.
\newblock Right practice, right place: {A} conservation planning toolbox for
  meeting water quality goals in the {Corn} {Belt}.
\newblock {\em Journal of Soil and Water Conservation}, 73(2):29A--34A, 2018.

\bibitem{schilling_modeling_2009}
Keith~E. Schilling and Calvin~F. Wolter.
\newblock Modeling {Nitrate}-{Nitrogen} {Load} {Reduction} {Strategies} for the
  {Des} {Moines} {River}, {Iowa} {Using} {SWAT}.
\newblock {\em Environmental Management}, 44(4):671--682, October 2009.

\bibitem{iowa2013science}
Iowa Nutrient~Reduction Strategy.
\newblock A science and technology based framework to assess and reduce
  nutrients to iowa waters and the gulf of mexico. prepared by iowa department
  of land stewardship, iowa department of natural resources and iowa state
  university college of agriculture and life sciences, 2013.

\bibitem{mclellan_reducing_2015}
E.~McLellan, D.~Robertson, K.~Schilling, M.~Tomer, J.~Kostel, D.~Smith, and
  K.~King.
\newblock Reducing {Nitrogen} {Export} from the {Corn} {Belt} to the {Gulf} of
  {Mexico}: {Agricultural} {Strategies} for {Remediating} {Hypoxia}.
\newblock {\em Journal of the American Water Resources Association},
  51(1):263--289, February 2015.

\bibitem{shortle_reforming_2012}
James~S. Shortle, Marc Ribaudo, Richard~D. Horan, and David Blandford.
\newblock Reforming {Agricultural} {Nonpoint} {Pollution} {Policy} in an
  {Increasingly} {Budget}-{Constrained} {Environment}.
\newblock {\em Environmental Science \& Technology}, 46(3):1316--1325, February
  2012.

\bibitem{mclellan2018right}
Eileen~L McLellan, Keith~E Schilling, Calvin~F Wolter, Mark~D Tomer, Sarah~A
  Porter, Joe~A Magner, Douglas~R Smith, and Linda~S Prokopy.
\newblock Right practice, right place: A conservation planning toolbox for
  meeting water quality goals in the corn belt.
\newblock {\em Journal of Soil and Water Conservation}, 73(2):29A--34A, 2018.

\bibitem{roy2021hot}
Eric~D Roy, Courtney R~Hammond Wagner, and Meredith~T Niles.
\newblock Hot spots of opportunity for improved cropland nitrogen management
  across the united states.
\newblock {\em Environmental Research Letters}, 16(3):035004, 2021.

\bibitem{olmstead_economics_2010}
Sheila~M. Olmstead.
\newblock The {Economics} of {Water} {Quality}.
\newblock {\em Review of Environmental Economics and Policy}, 4(1):44--62,
  January 2010.

\bibitem{laukkanen_evaluating_2014}
Marita Laukkanen and Céline Nauges.
\newblock Evaluating {Greening} {Farm} {Policies}: {A} {Structural} {Model} for
  {Assessing} {Agri}-environmental {Subsidies}.
\newblock {\em Land Economics}, 90(3):458--481, August 2014.

\bibitem{shortle_nutrient_2017}
James Shortle and Richard~D. Horan.
\newblock Nutrient pollution: {A} wicked challenge for economic instruments.
\newblock {\em Water Economics and Policy}, 3(02):1650033, 2017.

\bibitem{kucharik_evaluation_2003}
C.~J. Kucharik.
\newblock Evaluation of a {Process}-{Based} {Agro}-{Ecosystem} {Model}
  ({Agro}-{IBIS}) across the {US} {Corn} {Belt}: {Simulations} of the
  {Interannual} {Variability} in {Maize} {Yield}.
\newblock {\em Earth Interactions}, 7, 2003.

\bibitem{kucharik_integrated_2003}
C.~J. Kucharik and K.~R. Brye.
\newblock Integrated {BIosphere} {Simulator} ({IBIS}) yield and nitrate loss
  predictions for {Wisconsin} maize receiving varied amounts of nitrogen
  fertilizer.
\newblock {\em Journal of Environmental Quality}, 32(1):247--268, January 2003.

\bibitem{donner_corn-based_2008}
S.~D. Donner and C.~J. Kucharik.
\newblock Corn-based ethanol production compromises goal of reducing nitrogen
  export by the {Mississippi} {River}.
\newblock {\em Proceedings of the National Academy of Sciences of the United
  States of America}, 105(11):4513--4518, March 2008.

\bibitem{jaynes2010potential}
Dan~B Jaynes, Kelly~R Thorp, and David~E James.
\newblock Potential water quality impact of drainage water management in the
  midwest usa.
\newblock In {\em 9th International Drainage Symposium held jointly with CIGR
  and CSBE/SCGAB Proceedings, 13-16 June 2010, Qu{\~A}{\v{S}}bec City
  Convention Centre, Quebec City, Canada}, page~1. American Society of
  Agricultural and Biological Engineers, 2010.

\bibitem{cheng2020maximizing}
FY~Cheng, KJ~Van~Meter, DK~Byrnes, and NB~Basu.
\newblock Maximizing us nitrate removal through wetland protection and
  restoration.
\newblock {\em Nature}, 588(7839):625--630, 2020.

\bibitem{marshall2018reducing}
Elizabeth Marshall, Marcel Aillery, Marc Ribaudo, Nigel Key, Stacy Sneeringer,
  LeRoy Hansen, Scott Malcolm, and Anne Riddle.
\newblock Reducing nutrient losses from cropland in the mississippi/atchafalaya
  river basin: Cost efficiency and regional distribution.
\newblock Technical report, 2018.

\bibitem{christianson2013financial}
Laura Christianson, John Tyndall, and Matthew Helmers.
\newblock Financial comparison of seven nitrate reduction strategies for
  midwestern agricultural drainage.
\newblock {\em Water Resources and Economics}, 2:30--56, 2013.

\bibitem{valayamkunnath2020mapping}
Prasanth Valayamkunnath, Michael Barlage, Fei Chen, David~J Gochis, and
  Kristie~J Franz.
\newblock Mapping of 30-meter resolution tile-drained croplands using a
  geospatial modeling approach.
\newblock {\em Scientific data}, 7(1):1--10, 2020.

\bibitem{lima2019leakage}
Mairon G~Bastos Lima, U~Martin Persson, and Patrick Meyfroidt.
\newblock Leakage and boosting effects in environmental governance: a framework
  for analysis.
\newblock {\em Environmental Research Letters}, 14(10):105006, 2019.

\bibitem{dou2018spillover}
Yue Dou, Ramon Felipe~Bicudo da~SILVA, Hongbo Yang, and Jianguo Liu.
\newblock Spillover effect offsets the conservation effort in the amazon.
\newblock {\em Journal of Geographical Sciences}, 28(11):1715--1732, 2018.

\bibitem{fang2019clean}
Delin Fang, Bin Chen, Klaus Hubacek, Ruijing Ni, Lulu Chen, Kuishuang Feng, and
  Jintai Lin.
\newblock Clean air for some: Unintended spillover effects of regional air
  pollution policies.
\newblock {\em Science advances}, 5(4):eaav4707, 2019.

\bibitem{mclellan2015reducing}
Eileen McLellan, Dale Robertson, Keith Schilling, Mark Tomer, Jill Kostel, Doug
  Smith, and Kevin King.
\newblock Reducing nitrogen export from the corn belt to the gulf of mexico:
  Agricultural strategies for remediating hypoxia.
\newblock {\em JAWRA Journal of the American Water Resources Association},
  51(1):263--289, 2015.

\bibitem{fisher2004wetland}
J~Fisher and MC~Acreman.
\newblock Wetland nutrient removal: a review of the evidence.
\newblock {\em Hydrology and Earth system sciences}, 8(4):673--685, 2004.

\bibitem{prokopy2019adoption}
Linda~S Prokopy, Kristin Floress, J~Gordon Arbuckle, Sarah~P Church, Francis~R
  Eanes, Yuling Gao, Benjamin~M Gramig, Pranay Ranjan, and Ajay~S Singh.
\newblock Adoption of agricultural conservation practices in the united states:
  Evidence from 35 years of quantitative literature.
\newblock {\em Journal of Soil and Water Conservation}, 74(5):520--534, 2019.

\bibitem{masuda2021rented}
Yuta~J Masuda, Seth~C Harden, Pranay Ranjan, Chloe~B Wardropper, Collin Weigel,
  Paul~J Ferraro, Sheila~MW Reddy, and Linda~S Prokopy.
\newblock Rented farmland: A missing piece of the nutrient management puzzle in
  the upper mississippi river basin?
\newblock {\em Journal of Soil and Water Conservation}, 76(1):5A--9A, 2021.

\bibitem{niles2013perceptions}
Meredith~T Niles, Mark Lubell, and Van~R Haden.
\newblock Perceptions and responses to climate policy risks among california
  farmers.
\newblock {\em Global Environmental Change}, 23(6):1752--1760, 2013.

\bibitem{liu_achieving_2017}
Jing Liu, Thomas~W. Hertel, Richard~B. Lammers, Alexander Prusevich, Uris
  Lantz~C. Baldos, Danielle~S. Grogan, and Steve Frolking.
\newblock Achieving sustainable irrigation water withdrawals: global impacts on
  food security and land use.
\newblock {\em Environmental Research Letters}, 12(10):104009, 2017.
\newblock 00000.

\bibitem{baldos2020simple}
ULC Baldos, I~Haqiqi, TW~Hertel, Mark Horridge, and J~Liu.
\newblock Simple-g: A multiscale framework for integration of economic and
  biophysical determinants of sustainability.
\newblock {\em Environmental Modelling \& Software}, 133:104805, 2020.

\bibitem{kucharik_miscanthus_2013}
C.~J. Kucharik, A.~VanLoocke, J.~D. Lenters, and M.~M. Motew.
\newblock Miscanthus {Establishment} and {Overwintering} in the {Midwest}
  {USA}: {A} {Regional} {Modeling} {Study} of {Crop} {Residue} {Management} on
  {Critical} {Minimum} {Soil} {Temperatures}.
\newblock {\em Plos One}, 8(7), July 2013.

\bibitem{ramankutty_farming_2008}
Navin Ramankutty, Amato~T. Evan, Chad Monfreda, and Jonathan~A. Foley.
\newblock Farming the planet: 1. {Geographic} distribution of global
  agricultural lands in the year 2000.
\newblock {\em Global Biogeochemical Cycles}, 22(1):GB1003, March 2008.
\newblock 00457.

\bibitem{motew2013climate}
Melissa~M Motew and Christopher~J Kucharik.
\newblock Climate-induced changes in biome distribution, npp, and hydrology in
  the upper midwest us: A case study for potential vegetation.
\newblock {\em Journal of Geophysical Research: Biogeosciences},
  118(1):248--264, 2013.

\bibitem{TylerLark2021}
Tyler Lark, Nathan~P. Hendricks, Nicholas Pates, Aaron Smith, Seth~A. Spawn,
  Matthew Bougie, Eric~G. Booth, Christopher~J. Kucharik, and Holly~K. Gibbs.
\newblock Environmental outcomes from the u.s. renewable fuel standard.
\newblock {\em Proceedings of the National Academy of Sciences (in press)},
  2021.

\bibitem{sun2020fine}
Shanxia Sun, Brayam~Valqui Ordonez, Mort~D Webster, Jing Liu, Christopher~J
  Kucharik, and Thomas Hertel.
\newblock Fine-scale analysis of the energy--land--water nexus: Nitrate
  leaching implications of biomass cofiring in the midwestern united states.
\newblock {\em Environmental science \& technology}, 54(4):2122--2132, 2020.

\bibitem{ferguson2008neoclassical}
Charles~E Ferguson et~al.
\newblock {\em The neoclassical theory of production and distribution}.
\newblock Cambridge University Press, 2008.

\bibitem{mittergridded}
Soil Survey StaffSoil Survey StaffSoil Survey Staff.
\newblock {\em Gridded Soil Survey Geographic (gSSURGO-10) Database for the
  Conterminous United States-10 meter}. Available online at
  https://gdg.sc.egov.usda.gov/. April 17, 2017 (201701 official release).,
  April 2017.

\bibitem{USDANASS}
USDA-NASS.
\newblock {\em USDA National Agricultural Statistics Service Cropland Data
  Layer (2015). Available at https://nassgeodata.gmu.edu/CropScape/ (accessed
  04/02/17; verified 06/01/2021). Published crop-specific data layer [Online].}
\newblock USDA-NASS, Washington, DC., 2015.

\bibitem{Helmers2021}
M.~Helmers, L.~Abendroth, L.B. Reinhart, G.~Chighladze, L.~Pease, L.~Bowling,
  M.~Y.~Ehsan Ghane, L.~Ahiablame, L.~Brown, N.~Fausey, J.~Frankenberger,
  D.~Jaynes, K.~King, E.~Kladivko, K.~Nelson, and J.~Strock.
\newblock Controlled drainage impacts on subsurface drain flow and nitrate
  load: A synthesis of studies across the u.s. midwest and southeast.
\newblock 2021.

\bibitem{lee2017regional}
{\em Regional variability of subsurface drainage in the {US} corn belt}, 2017.

\bibitem{crumpton2001using}
WG~Crumpton.
\newblock Using wetlands for water quality improvement in agricultural
  watersheds; the importance of a watershed scale approach.
\newblock {\em Water Science and Technology}, 44(11-12):559--564, 2001.

\bibitem{Ward2016}
A.D. Ward, S.R. Trimble, S.W.~and Burckhard, and J.G. Lyon.
\newblock {\em Environmental Hydrology}.
\newblock CRC Press, Boca Raton, FL, USA , 2016.

\bibitem{tchobanoglous1985water}
George Tchobanoglous and Edward~E Schroeder.
\newblock {\em Water quality: characteristics, modeling, modification}.
\newblock Addison-Wesley Pub. Co., Reading, MA, 1985.

\bibitem{kadlec2009treatment}
RH~Kadlec and SD~Wallace.
\newblock Treatment wetlands 2nd edition crc press, 2009.

\bibitem{mitsch_wetland_2009}
William~J. Mitsch, James~G. Gosselink, Li~Zhang, and Christopher~J. Anderson.
\newblock {\em Wetland ecosystems}.
\newblock John Wiley \& Sons, 2009.

\bibitem{Crumpton2006}
William~G Crumpton, Greg~A Stenback, Bradley~Allen Miller, and Matthew~J
  Helmers.
\newblock Potential benefits of wetland filters for tile drainage systems:
  impact on nitrate loads to mississippi river subbasins.
\newblock Technical report, AGRICULTURAL AND BIOSYSTEMS ENGINEERING TECHNICAL
  REPORTS AND WHITE PAPERS, 2006.

\bibitem{davis1981prairie}
B.~Richardson, editor.
\newblock {\em Prairie pothole marshes as traps for nitrogen and phosphorous in
  agricultural runoff}, The Freshwater Society, MN., June 1981. Selected
  Proceedings of the Midwest Conference on Wetland Values and Management.

\bibitem{kovacic2000effectiveness}
David~A Kovacic, Mark~B David, Lowell~E Gentry, Karen~M Starks, and Richard~A
  Cooke.
\newblock Effectiveness of constructed wetlands in reducing nitrogen and
  phosphorus export from agricultural tile drainage.
\newblock Technical report, Wiley Online Library, 2000.

\bibitem{phipps1994factors}
Richard~G Phipps and William~G Crumpton.
\newblock Factors affecting nitrogen loss in experimental wetlands with
  different hydrologic loads.
\newblock {\em Ecological Engineering}, 3(4):399--408, 1994.

\bibitem{phipps1997nitrate}
Richard~Gregory Phipps.
\newblock {\em Nitrate removal capacity of constructed wetlands}.
\newblock Iowa State University, 1997.

\bibitem{mitsch2005nitrate}
William~J Mitsch, John~W Day, Li~Zhang, and Robert~R Lane.
\newblock Nitrate-nitrogen retention in wetlands in the mississippi river
  basin.
\newblock {\em Ecological engineering}, 24(4):267--278, 2005.

\bibitem{zhang2000hydrologic}
W.~J. Mitsch and L.~Zhang, editors.
\newblock {\em Hydrologic budgets of the two Olentangy River experimental
  wetlands, 1994-99}, Annual Report 1999:41-46., 2000. Olentangy River Wetland
  Research Park at the Ohio State University.

\bibitem{zhang2000water}
W.~J. Mitsch and L.~Zhang, editors.
\newblock {\em Water budgets of the two Olentangy River experimental wetlands
  in 2000}, Annual Report 2000:17-28., 2001. Olentangy River Wetland Research
  Park at the Ohio State University.

\bibitem{Zhang2002}
W.~J. Mitsch and L.~Zhang, editors.
\newblock {\em Water budgets of the two Olentangy River experimental wetlands
  in 2001}, Annual Report 2001: 23-34., 2002. Olentangy River Wetland Research
  Park at the Ohio State University.

\bibitem{zhang2004water}
W.~J. Mitsch, L.~Zhang, and C.~Tuttle, editors.
\newblock {\em Water budgets of the two Olentangy River experimental wetlands
  in 2003}, Annual Report 2003, pp. 39-52., 2004. Olentangy River Wetland
  Research Park at the Ohio State University.

\bibitem{maurer2002long}
Edwin~P Maurer, Andrew~W Wood, Jennifer~C Adam, Dennis~P Lettenmaier, and Bart
  Nijssen.
\newblock A long-term hydrologically based dataset of land surface fluxes and
  states for the conterminous united states.
\newblock {\em Journal of climate}, 15(22):3237--3251, 2002.

\end{thebibliography}
\end{document}